\documentstyle[12pt]{article}
\catcode`\@=11

\@addtoreset{equation}{section}

\def\@normalsize{\@setsize\normalsize{15pt}\xiipt\@xiipt
\abovedisplayskip 14pt plus3pt minus3pt%
\belowdisplayskip \abovedisplayskip
\abovedisplayshortskip  \z@ plus3pt%
\belowdisplayshortskip  7pt plus3.5pt minus0pt}
\def\small{\@setsize\small{13.6pt}\xipt\@xipt
\abovedisplayskip 13pt plus3pt minus3pt%
\belowdisplayskip \abovedisplayskip
\abovedisplayshortskip  \z@ plus3pt%
\belowdisplayshortskip  7pt plus3.5pt minus0pt
\def\@listi{\parsep 4.5pt plus 2pt minus 1pt
            \itemsep \parsep
            \topsep 9pt plus 3pt minus 3pt}}

\def\underline#1{\relax\ifmmode\@@underline#1\else
        $\@@underline{\hbox{#1}}$\relax\fi}
\@twosidetrue
\relax

\catcode`@=12

\catcode`\@=11

\def\section{\@startsection{section}{1}{\z@}{3.5ex plus 1ex minus
   .2ex}{2.3ex plus .2ex}{\large\bf}}

\def\ps@headings{\def\@oddfoot{}\def\@evenfoot{}
\def\@oddhead{\hbox{}\hfill
        \makebox[.5\textwidth]{\raggedright\ignorespaces --\thepage{}--
        \hfill }}
\def\@evenhead{\@oddhead}
\def\subsectionmark##1{\markboth{##1}{}}
}

\ps@headings

\catcode`\@=12

\relax

\def\figcap{\section*{Figure Captions\markboth
        {FIGURECAPTIONS}{FIGURECAPTIONS}}\list
        {Fig. \arabic{enumi}:\hfill}{\settowidth\labelwidth{Fig. 999:}
        \leftmargin\labelwidth
        \advance\leftmargin\labelsep\usecounter{enumi}}}
 \relax
\def\tablecap{\section*{Table Captions\markboth
        {TABLECAPTIONS}{TABLECAPTIONS}}\list
        {Table \arabic{enumi}:\hfill}{\settowidth\labelwidth{Table 999:}
        \leftmargin\labelwidth
        \advance\leftmargin\labelsep\usecounter{enumi}}}
 \relax
\def\reflist{\section*{References\markboth
        {REFLIST}{REFLIST}}\list
        {[\arabic{enumi}]\hfill}{\settowidth\labelwidth{[999]}
        \leftmargin\labelwidth
        \advance\leftmargin\labelsep\usecounter{enumi}}}
 \relax

\catcode`\@=11

\def\ps@headings{\def\@oddfoot{}\def\@evenfoot{}
\def\@oddhead{\hbox{}\hfill
        \makebox[.5\textwidth]{\raggedright\ignorespaces --\thepage{}--
        \hfill }}
\def\@evenhead{\@oddhead}
\def\subsectionmark##1{\markboth{##1}{}}
}

\ps@headings

\relax

\def\firstpage#1#2#3#4#5#6{
\begin{document}
\begin{titlepage}
\nopagebreak
\title{\begin{flushright}
        \vspace*{-1.8in}
%        {\normalsize hep-th/9512006}\\[-9mm]
%       {\normalsize NUB--#1 -- #2\\[-9mm]
   {\normalsize CPTH--S397.1195 -- LPTENS--95/55}\\[-9mm]
        {\normalsize hep-th/9512006}\\[4mm]
\end{flushright}
\vfill
{#3}}
\author{\large #4 \\[1.0cm] #5}
\maketitle
\vskip -7mm
\nopagebreak
\begin{abstract}
{\noindent #6}
\end{abstract}
\vfill
\begin{flushleft}
\rule{16.1cm}{0.2mm}\\[-3mm]
$^{\star}${\small Research supported in part by\vspace{-4mm}
the National Science Foundation under grant
PHY--93--06906, \linebreak in part by the EEC contracts \vspace{-4mm}
SC1--CT92--0792 and CHRX-CT93-0340,  and in part by
CNRS--NSF  grant INT--92--16146.}\\[-3mm]
%\vspace{-4mm}
%and in part by CNRS--NSF
%grant INT--92--16146.}\\[-3mm]
$^{\dagger}${\small Laboratoire Propre du CNRS UPR A.0014.}\\[-3mm]
$^{\natural}${\small Unit\'e Propre du CNRS, associ\'ee \`a l'Ecole Normale
Sup\'erieure et \`a l'Universit\'e de Paris-Sud.}
December 1995
\end{flushleft}
\thispagestyle{empty}
\end{titlepage}}
\newcommand{\dal}{\raisebox{0.085cm}
{\fbox{\rule{0cm}{0.07cm}\,}}}
\newcommand{\dt}{\partial_{\langle T\rangle}}
\newcommand{\dtbar}{\partial_{\langle\bar{T}\rangle}}
\newcommand{\al}{\alpha^{\prime}}
\newcommand{\mst}{M_{\scriptscriptstyle \!S}}
\newcommand{\mpl}{M_{\scriptscriptstyle \!P}}
\newcommand{\dv}{\int{\rm d}^4x\sqrt{g}}
\newcommand{\lv}{\left\langle}
\newcommand{\rv}{\right\rangle}
\newcommand{\ph}{\varphi}
\newcommand{\abar}{\bar{a}}
\newcommand{\sbar}{\,\bar{\! S}}
\newcommand{\xbar}{\,\bar{\! X}}
\newcommand{\fbar}{\,\bar{\! F}}
\newcommand{\zbar}{\bar{z}}
\newcommand{\dbar}{\,\bar{\!\partial}}
\newcommand{\tbar}{\bar{T}}
\newcommand{\taubar}{\bar{\tau}}
\newcommand{\ubar}{\bar{U}}
\newcommand{\ybar}{\bar{Y}}
\newcommand{\phb}{\bar{\varphi}}
\newcommand{\cm}{Commun.\ Math.\ Phys.~}
\newcommand{\pr}{Phys.\ Rev.\ D~}
\newcommand{\pl}{Phys.\ Lett.\ B~}
\newcommand{\ibar}{\bar{\imath}}
\newcommand{\jbar}{\bar{\jmath}}
\newcommand{\np}{Nucl.\ Phys.\ B~}
\newcommand{\F}{{\cal F}}
\renewcommand{\L}{{\cal L}}
\newcommand{\A}{{\cal A}}
\newcommand{\e}{{\rm e}}
\newcommand{\be}{\begin{equation}}
\newcommand{\en}{\end{equation}}
\newcommand{\gsi}{\,\raisebox{-0.13cm}{$\stackrel{\textstyle
>}{\textstyle\sim}$}\,}
\newcommand{\lsi}{\,\raisebox{-0.13cm}{$\stackrel{\textstyle
<}{\textstyle\sim}$}\,}
\date{}
\firstpage{3118}{IC/95/34}
{\large\bf Spontaneous Breaking of $N=2$ Global Supersymmetry$^{\star}$}
{I. Antoniadis$^{\,a}$,  H. Partouche$^{\,a}$
and T.R. Taylor$^{\,b,c}$}%\\[-3mm]
{\normalsize\sl
$^a$Centre de Physique Th\'eorique, Ecole Polytechnique,$^\dagger$
{}F-91128 Palaiseau, France\\[-3mm]
\normalsize\sl
$^b$Laboratoire de Physique Th\'eorique de l'Ecole Normale
Sup\'erieure,$^\natural$\\[-5mm]\normalsize\sl
24, rue Lhomond, F-75231 Paris, France\\[-3mm]
\normalsize\sl $^c$Department of Physics, Northeastern
University, Boston, MA 02115, U.S.A.}
{We study spontaneous supersymmetry breaking in $N=2$ globally supersymmetric
theories describing a system of abelian vector multiplets. We find that the
most general form of the action admits, in addition to the usual
Fayet-Iliopoulos term, a magnetic Fayet-Iliopoulos term for the auxiliary
components of dual vector multiplets. In a generic case, $N=2$ supersymmetry
is broken down spontaneously to $N=1$. In some cases however, the scalar
potential can drive the theory towards a $N=2$ supersymmetric ground state
where massless dyons condense in the vacuum.}

The framework of globally supersymmetric field theories is highly restrictive,
allowing very few mechanisms for spontaneous supersymmetry breaking.
In $N=1$ supersymmetric theories there are basically two types of
breaking: F-type and D-type,  with the auxiliary components
of chiral and  vector superfields, respectively, acquiring non-vanishing
vacuum expectation values (VEVs). F-type breaking is usually induced by a
non-trivial superpotential while D-type breaking is generically due to the
presence of a Fayet-Iliopoulos (FI) term associated with a $U(1)$ factor in the
gauge group.\footnote{For a recent review, see \cite{t}.}

$N=2$ supersymmetry is even more restrictive, with only one
mechanism known up to date to break it, based on $N=1$ Fayet-Iliopoulos term
\cite{fay}. It can be realized  in the presence of a $N=2$ vector multiplet
associated to an abelian gauge group factor. Decomposed under $N=1$
supersymmetry, such a  multiplet contains one vector and one chiral multiplet.
A FI term is also equivalent to a superpotential
which is linear in the chiral superfield. No other superpotential
seemed to be allowed for chiral components of $N=2$ vector multiplets.

Recently there has been revived interest in $N=2$ supersymmetry, in particular
in the effective actions describing non-perturbative dynamics of non-abelian
gauge  theories. In general, these theories exist only in the Coulomb phase,
with a number of abelian vector multiplets and possibly hypermultiplets,
and their low energy effective actions can be determined exactly by using
the underlying duality symmetries \cite{sw}.

In this work, we study $N=2$ supersymmetric actions describing a system of
abelian vector multiplets. Since we are interested in these theories
viewed as low-energy realizations of some more complicated
physical systems, we do not impose the renormalizability requirement and
consider the most general form of the Lagrangian. We find that
$N=2$ supersymmetry admits also another form of the superpotential
which can be interpreted as a Fayet-Iliopoulos term for a  ``magnetic''
$U(1)$. A non-trivial potential can then be generated for the
scalar fields. As a result, we find a novel mechanism for $N=2$
supersymmetry breaking. Even more surprisingly, we find that
$N=2$ supersymmetry can be broken {\it partially\/} to $N=1$ already
at the global level.\footnote{It has been shown recently that partial breaking
may occur in the framework of {\it local\/} supersymmetry \cite{f}.}

The basic points of our analysis can be explained on the simplest example
of $N=2$ supersymmetric gauge theory with one abelian vector multiplet $A$
which contains besides the $N=1$ gauge multiplet $(\A_{\mu},\lambda)$
a neutral chiral superfield $(a,\chi)$.
For the sake of clarity, we begin with $N=1$ superfield description and
rederive our results later on by using the full $N=2$ formalism.
In the absence of superpotential and FI term, the most
general Lagrangian describing this theory is determined by the analytic
prepotential
$\F(A)$, in terms of which the K\"ahler potential $K$ and the gauge
kinetic function $f$ are given by:
\be
K(a,\abar)={i\over 2}(a{\bar\F}_{\abar}-\abar \F_a)\qquad\qquad
f(a)=-i\F_{aa}\ ,
\label{Kf}
\en
where the $a$ and $\abar$ subscripts denote derivatives with respect to $a$ and
$\abar$, respectively. In $N=1$ superspace, the Lagrangian is written as:
\be
{\cal L}_0=\frac{1}{4}\int d^2\theta f{\cal W}^2+c.c. +\int d^2\theta
d^2{\bar\theta}K
\label{L0}
\en
where $\cal W$ is the standard gauge field strength superfield.\footnote{We use
the conventions of ref.\cite{wb}.}

The Lagrangian ${\cal L}_0$ can be supplemented by a FI term
which is linear in the auxiliary $D$ component of the gauge vector multiplet:
\be
\L_D=\sqrt{2}\xi D\ , \label{LD}
\en
with $\xi$ a real constant. It is well known that such a term preserves
also $N=2$ supersymmetry \cite{fay}.

The Lagrangian ${\cal L}_0$ can also be supplemented by a superpotential term
\be
\L_W=\int d^2\theta\, W+c.c.
\label{LW}
\en
In order to determine what form of the superpotential
is compatible with $N=2$ supersymmetry we will impose the
constraint that the full Lagrangian,
\be
\L=\L_0+\L_D+\L_W\ ,
\label{L}
\en
be invariant under the exchange of the gaugino $\lambda$ with the fermion
$\chi$. This condition is necessary for the global $SU(2)$ symmetry under which
$(\chi,\lambda)\equiv(\lambda_1,\lambda_2)$ transforms as a doublet. It is
easy to see that it is satisfied provided that the $\lambda\lambda$ and
$\chi\chi$ mass terms are equal. It follows that
\be
W_{aa}+i{\tau_a\over 2\tau_2}W_a=
i{\tau_a\over 2\tau_2}{\overline W}_{\abar}\ ,
\label{masses}
\en
where we defined $\tau\equiv \F_{aa}=\tau_1+i\tau_2$. The left- and the
right-hand sides of the above equation correspond to the $\chi$ and
$\lambda$ mass terms, respectively. Its general solution is:
\be
W=ea+m\F_a\ ,\label{W}
\en
up to an irrelevant additive constant. Here $e$ and $m$ are arbitrary real
numbers. For $m=0$ the above superpotential is equivalent to a FI
term (\ref{LD}) with $\xi=e$ \cite{fay}.

After eliminating the auxiliary fields, $\L_D+\L_W$ gives rise to only two
modifications in the original Lagrangian $\L_0$. It induces the fermion mass
terms mentioned before, $\frac{1}{2}{\cal M}_{ij}\lambda_i\lambda_j$, with
\be
{\cal M}={i\over 2}\tau_a\pmatrix{e+m\taubar&i\xi\cr
i\xi&e+m\taubar}
\label{fmasses}
\en
and the scalar potential
\be
V_{N{=}1}={|e+m\tau|^2+\xi^2\over\tau_2}\ .
\label{V1}
\en

In order to prove that the full Lagrangian (\ref{L}) is indeed invariant
under $N=2$ supersymmetry, we will rederive it by using the $N=2$ superspace
formalism. In this formalism, $N=2$  vector multiplets are described
by reduced chiral superfields. The reducing constraint \cite{red}
\be
(\epsilon_{ij}D^i\sigma_{\mu\nu}D^j)^2A=-96\Box A^*
\label{red}
\en
eliminates unwanted
degrees of freedom, in particular by imposing the Bianchi identity for the
gauge
field strength. In terms of the reduced chiral superfield $A$, the
Lagrangian $\L_0$ can then be written as
${i\over 4}\int d^2\theta_1 d^2\theta_2 \F(A) + c.c.$
$\L_0$ can also be written in terms of an {\it unconstrained} superfield $A$
as:
\be
\L_0={i\over 4}\int d^2\theta_1 d^2\theta_2 [\F(A) - A_D A] + c.c.
\label{L02}
\en
where $A_D$ is reduced superfield which plays the role of a Lagrange
multiplier. Its equation of motion imposes the reducing constraint on $A$.
$A_D$ can also be identified with the dual superfield; the standard
duality transformation amounts to rewriting $\L_0$ in terms of $A_D$ after
eliminating $A$ with the use of its equations of motion, $A_D=\F_A$.

As in the $N=1$ case $\L_0$ can be supplemented with a Fayet-Iliopoulos term,
linear in the auxiliary components of $A$. $N=2$ auxiliary fields form an
$SU(2)$ triplet $\vec Y$ with components $Y_n$, $n=1,2,3$. For a reduced
superfield, $Y_n$ are real and can be identified with the $N=1$ auxiliary
components $F$ and $D$ as follows:
\be
Y_1+iY_2=2iF\qquad\qquad Y_3={\sqrt 2}D\ .\label{aux}
\en
Under $N=2$ supersymmetry transformations, these auxiliary fields transform
into total derivatives. Hence, a term linear in $\vec Y$ can be added to the
action:
\be
\L_D={1\over 2} {\vec E}\cdot{\vec Y} + c.c.\ ,\label{LD2}
\en
where $E_n$ are arbitrary parameters which can be chosen to
be real since their imaginary parts drop from the action. Furthermore, using
the global $SU(2)$ symmetry one can choose $\vec E$ to point in any direction.
For instance, by choosing $E_1=E_2=0$, $\L_D$ becomes equivalent to (\ref{LD})
with $\xi =E_3$, while for $E_1=E_3=0$, $\L_D$ is equivalent to a $N=1$
superpotential (\ref{W}) with $e=E_2$ and $m=0$.

A FI term can also be introduced for the auxiliary component ${\vec Y}_D$ of
the dual superfield $A_D$, see eq.(\ref{L02}):
\be
\L_D'={1\over 2} {\vec M}\cdot{\vec Y}_D + c.c.\ ,\label{LDD}
\en
with arbitrary real $M_n$.  The full Lagrangian becomes:
\be
\L=\L_0+\L_D+\L_D'=
{i\over 4}\int d^2\theta_1 d^2\theta_2 [\F(A) - A_D A] +
{1\over 2} ({\vec E}\cdot{\vec Y} + {\vec M}\cdot{\vec Y}_D) + c.c.
\label{L2}
\en
Note that the presence of $\L_D'$ affects the equation of motion with
respect to the auxiliary component of the Lagrange multiplier ${\vec Y}_D$. As
a
result, the auxiliary component
$\vec Y$ is no longer constrained to be real and acquires a constant
imaginary part,
$\makebox{Im}{\vec Y}=2{\vec M}$. Hence the full action depends on
both the real and the imaginary part of $\vec E$. Together with
the real $\vec M$ we have now 9 real parameters.

In order to make contact
with the $N=1$ Lagrangian (\ref{L}), we perform an $SU(2)$ transformation
which brings the parameters $\vec M$ and Re$\vec E$ into the form
\be
\vec{M}=\pmatrix{0& m &0}\qquad\qquad
\makebox{Re}\vec{E}=\pmatrix{0& e&\xi}\ . \label{su2}
\en
It is now straightforward to show that after elimination of auxiliary fields
the $N=2$ Lagrangian (\ref{L2}) coincides with (\ref{L}) up to an additive
field-independent constant. Indeed, the scalar potential is given by:
\be
V ~=~ {|\makebox{Re}{\vec E}+{\vec M}\tau |^2\over\tau_2}+
2{\vec M}\cdot\makebox{Im}{\vec E} ~=~ V_{N{=}1}+2mp\ ,
\label{V}
\en
where $p\equiv\makebox{Im}E_2$.

{}From the above discussion it is clear that the D-term (\ref{LDD})
involving the parameter $\vec M$ corresponds to a Fayet-Iliopoulos
term for the dual magnetic $U(1)$ gauge field. It can be obtained
from the standard electric D-term (\ref{LD2}) by a duality
transformation $A\rightarrow A_D$. Indeed, by performing
a symplectic $Sp(2,R)\simeq SL(2,R)$ change of basis
\be
\pmatrix{\F_a\cr a}\to\pmatrix{\alpha&\beta\cr\gamma&\delta}\pmatrix{\F_a\cr a}
\quad\qquad \tau\to{\alpha\tau+\beta\over\gamma\tau+\delta}       \label{symp}
\en
with $\alpha\delta-\beta\gamma=1$,  one obtains from (\ref{L2})
the same form of Lagrangian with new parameters $\vec M'$ and
$\vec E'$ given by
\be
\pmatrix{\vec{M}'& \makebox{Re}\vec{E}'}=
\pmatrix{\vec{M}& \makebox{Re}\vec{E}}
\pmatrix{\alpha&\beta\cr\gamma&\delta}
\qquad\quad \makebox{Im}\vec{E}'={\vec{M}\cdot
\makebox{Im}\vec{E}\over M'^2}\,\vec{M}' \ .                 \label{prime}
\en
In the $SU(2)$ ``gauge'' of eq.(\ref{su2}), this corresponds to
\be
m'=[(\alpha m+\gamma e)^2+\gamma^2\xi^2]^{1/2}
\quad e'={(\alpha m+\gamma e)(\beta m+\delta e)+\gamma\delta
\xi^2\over m'}\quad \xi'=\frac{\xi m}{m'}\quad p'={pm\over m'}\ .
\label{prime2}
\en

We now turn to the minimization of the scalar potential (\ref{V}).
It has obviously a stationary point at $\tau_a=0$.
However, this is in general a saddle point since $V_{a\abar}=0$.\footnote{
A possible exception could arise in an unlikely case of an
essential zero of $\tau_a$ at this point.}

For $m\ne 0$, a stable minimum exists at\footnote{Without losing
generality we can choose $m,\xi\ge 0$.}
\be
\tau_1=-{e\over m}\qquad\qquad \tau_2={\xi\over m}\ ,
\label{min}
\en
with the  minimum value $V=2m(\xi+p)$. In this vacuum, the complex scalar
$a$ acquires the mass ${\cal M}_a=m|\tau_a|$. After diagonalizing the fermion
mass matrix (\ref{fmasses})  we find one massless fermion $(\chi-\lambda)
/\sqrt{2}$, and one massive spinor $(\chi+\lambda)/\sqrt{2}$, with the Majorana
mass ${\cal M}_a$ equal to the scalar mass. This degeneracy is not accidental.
As we explain below, the vacuum (\ref{min}) preserves $N=1$ supersymmetry,
and the spectrum consists of one massless vector and one massive
chiral multiplets.

In order to discuss supersymmetry breaking, it is sufficient to examine the
auxiliary field dependence of fermion transformations under $N=2$
supersymmetry:
\be
\delta\lambda_i=\frac{i}{\sqrt 2}Y_n\epsilon_{ij}(\sigma^n)^j_k\eta^k +\dots
\label{tr}
\en
where $\sigma^n$ are the Pauli matrices and the spinors $\eta^k$, $k=1,2$, are
the transformation parameters. As we have shown before, the effect of
the magnetic FI term (\ref{LDD}), after elimination of the Lagrange
multiplier, amounts to introducing a constant imaginary part for $\vec Y$. In
fact, the reducing constraint for $N=2$ chiral superfields (\ref{red})
implies $\Box{\vec Y}=\Box{\vec Y}^*$, which leaves precisely the
same freedom. This constant, $\makebox{Im}{\vec Y}=2{\vec M}$, enters into the
supersymmetry transformations (\ref{tr}) implying that generically both
supersymmetries are realized in a spontaneously broken mode. However, at the
minimum (\ref{min}) the real part of $\vec Y$ acquires an expectation value, so
that:
\be
{\vec Y} ~=~ -{2\over\tau_2}(\makebox{Re}{\vec E}+{\vec M}\tau_1)
+2i{\vec M} ~=~ 2m\pmatrix{0&i&-1}
\label{Ymin}
\en
As a result,
\be
\delta{\chi+\lambda\over \sqrt 2}=0\qquad\qquad
\delta{\chi-\lambda\over \sqrt 2}=-2im(\eta^1-\eta^2)
\label{deltas}
\en
which shows that one supersymmetry, corresponding to the diagonal
combination of the two, is preserved while the other one is spontaneously
broken. The
massless goldstino is identified as $(\chi-\lambda)/{\sqrt 2}$, in agreement
with the spectrum found before.

The presence of the magnetic FI term in (\ref{L2}) introduces into
the Lagrangian one additional parameter besides $m$, $p=\makebox{Im}E_2$,
which enters only as an additive constant in the scalar potential (\ref{V}).
{}From the above analysis it is natural to choose $p=-\xi$ so that
the potential vanishes at the $N=1$ supersymmetric minimum (\ref{min}).
Note that this is consistent with the symplectic transformations
(\ref{prime2}).
The ``cosmological'' constant $p$ could play important role once
the theory is coupled to gravity.

For $\xi=0$ the minimum (\ref{min}) occurs at a point where the metric
$\tau_2$ vanishes. This can happen either at ``infinity'' of the $a$-space
or at finite singular points where massless particles appear.
The quantum numbers of such states, including electric and magnetic charges,
as well as quantization conditions, depend on details of the underlying theory.
Its dynamics determines also the non-perturbative symmetries
which form a (discrete) subgroup of $Sp(2,R)$. These states cannot be vector
multiplets since unbroken non-abelian gauge group is incompatible with FI
terms.
Hence we assume that they
are BPS-like dyons which form $N=2$ hypermultiplets and that the minimization
condition (\ref{min}) defines a point $a=a_0$ where one of these
hypermultiplets
becomes massless. This can happen only if the parameters $(m,e)$ are
proportional to its magnetic and electric charges $(m_0,e_0)$, $(m,e)=c
(m_0,e_0)$. In order to analyze the behavior of the theory near $a_0$, one has
to include the massless hypermultiplet in the effective field theory
as a new degree of freedom. This can be done by performing the
duality transformation $A\to{\tilde A}=e_0A+m_0\F_A$, which makes possible
local
description of the dyon-gauge boson interactions. In $N=1$ superspace the
superpotential (\ref{W}) becomes:
\be
W=c{\tilde a} + {\sqrt 2}{\tilde a}{\phi^+}\phi^- \ ,
\label{W0}
\en
where $\phi^{\pm}$ are the two chiral superfield components of the
hypermultiplet, and ${\tilde a}$ is the chiral component of ${\tilde A}$.

The superpotential (\ref{W0}) describes $N=2$ QED with a Fayet-Iliopoulos
term proportional to $c$ \cite{fay}.
The minimization conditions of the respective potential are
$W_{\phi^{\pm}}=0$ which is automatically satisfied at ${\tilde a}=0$ ($a=a_0$)
and
\be
W_{\tilde a}=c+{\sqrt 2}{\phi^+}\phi^-=0\ ,\qquad
{\tilde D}=0=|\phi^+|^2-|\phi^-|^2\ .
\label{Wa}
\en
As a result the dyonic hypermultiplet condenses in a $N=2$ supersymmetric
vacuum. For instance if $e=0$, the dyonic state is a pure monopole and the VEV
of the scalar field $a$ is driven to the point where the monopole becomes
massless and acquires a non-vanishing expectation value. Its condensation
breaks
the magnetic $U(1)$ and imposes confinement of electric charges. This situation
is similar to the case considered in ref.\cite{sw} in the context of $SU(2)$
Yang-Mills with an explicit mass term for chiral components of gauge
multiplets which breaks $N=2$ supersymmetry explicitly to $N=1$.

For $m=0$, the scalar potential (\ref{V}) has a runaway behavior, $V\to 0$ as
$\tau_2\to\infty$. This case is equivalent by a duality transformation to the
the case $m\neq 0$, $\xi=0$ discussed above. The runaway behavior can
be avoided if there are singular points corresponding to massless electrically
charged particles. At these points the metric
$\tau_2$ has a logarithmic singularity and the massless states have to be
included explicitly in the low energy Lagrangian to avoid non-localities. A
similar analysis of the effective theory shows that $a$ is driven then to the
points where the massless hypermultiplets get non-vanishing VEVs breaking the
$U(1)$ gauge symmetry while $N=2$ supersymmetry remains unbroken.

In the context of string theory, this phenomenon is similar to the effect
induced by a generic superpotential near the conifold singularity of
type II superstrings compactified on a Calabi-Yau manifold \cite{s}.
In this case, the massless hypermultiplets are
black holes which condense at the conifold points. It has been shown
recently that such a superpotential can be generated by a VEV of the
10-form which in four dimensions corresponds to a magnetic FI term,
and that the black hole condensation at the conifold point leads
to new $N=2$ type II superstring vacua  \cite{ps}.

The generalization of the above analysis to the case of several abelian
multiplets is straightforward. In addition to the usual ``electric'' FI term
for a linear combination of D-terms one can add a ``magnetic" FI term
for a different combination. It is clear from our discussion in the framework
of $N=1$ supersymmetry that in a general $N=2$ supersymmetric theory one can
introduce three parameters $m,e,\xi$ for each abelian vector multiplet. It is a
very interesting question whether electric and magnetic Fayet-Iliopoulos terms
described in this work can be generated dynamically, for instance by an
underlying non-abelian gauge theory. It is clear that instantons do not
generate them since they give rise only to correlation functions
involving at least four fermions \cite{sei} whereas FI terms are associated
with fermion bilinears (\ref{fmasses}). However, one cannot a priori
exclude the existence of other non-perturbative effects, possibly related to
gaugino condensation, which could generate this type of terms in the effective
action.

\noindent{\bf Acknowledgments}

We thank E. Cremmer, P. Fayet, S. Ferrara and J. Iliopoulos for useful
conversations. I.A. acknowledges the Department of Physics at Northeastern
University for its kind hospitality.

\end{document}